\documentclass[twocolumn,showpacs,pre]{revtex4}

\usepackage{graphicx}
\usepackage{dcolumn}
\usepackage{bm}
\begin{document}

\title{Thermodynamic Rule Determining the Biological DNA Information Capacity}
\author{A. Widom and J. Swain}
\affiliation{Physics Department, Northeastern University, Boston MA USA}
\author{Y.N. Srivastava}
\affiliation{Physics Department \& INFN, University of Perugia, Perugia IT}
\author{S. Sivasubramanian}
\affiliation{Center for High-Rate Nanomanufacturing, Northeastern University, Boston, MA USA}
\author{V.I. Valenzi}
\affiliation{Centro Studi di Biometeorologia Onlus Roma/Lugano Via Besso 59, Lugano CH}

\begin{abstract}
A rigorous thermodynamic expression is derived for the total biological 
information capacity per unit length of a DNA molecule. The total 
information includes the usual four letter coding sequence information plus 
that excess information coding often erroneously referred to as ``junk''. 
We conclude that the currently understood human DNA code is about a hundred megabyte 
program written on a molecule with about a ten gigabyte memory. By far, most 
of the programing code is not presently understood. 
\end{abstract}


\maketitle

\section{Introduction \label{intro}}

The information capacity \begin{math} {\cal H} \end{math} in a 
human DNA molecule of length 
\begin{math} L\sim 3\ {\rm meter} \end{math} arising from the 
conventional four letter sequence code has been estimated 
to be\cite{Percus:2002} 
\begin{equation}
{\cal H}_{\rm code} \sim  10^{9}\ {\rm bit}.
\label{intro1}
\end{equation}
This estimate is approximately correct for humans, apes and perhaps snails. 
While the authors might admit some similarities with the apes, they 
would object to being called similar to snails. In defense of 
claiming that humans are higher on the evolutionary scale, one might 
include the so-called ``junk'' DNA residing in the chain. The estimates 
are 
\begin{equation}
{\cal H}_{\rm code+junk} \sim  10^{11}\ {\rm bit}.
\label{intro2}
\end{equation}
In terms of the total information capacity, humans do indeed 
appear to be on a higher evolutionary scale than (say) snails. 

There has been considerable recent interest in the nature of ``junk'' DNA 
sequences\cite{Willingham:2006,Riley:2002,Makalowski:2003,Gregory:1999} 
and in particular the role that they play in the evolutionary process. 
Our purpose is to derive a thermodynamic expression for the information 
which resides in DNA. In the thermodynamic limit, the 
information capacity per unit length  
\begin{equation}
\eta =\lim_{L\to \infty} \frac{\cal H}{L}
\label{intro3}
\end{equation}
has a thermodynamic description which is both mathematically rigorous and 
yet is experimentally measurable. The rule may be stated in terms of the DNA 
chain tension \begin{math} \tau  \end{math} as a function of temperature 
\begin{math} T \end{math} and the chemical potentials 
\begin{math} (\mu_1, \mu_2, \ldots ,\mu_c) \end{math} of the molecules 
which make up the DNA chain; i.e. 
\begin{equation}
\tau=\tau (T,\mu_1, \mu_2, \ldots ,\mu_c)
\label{intro4}
\end{equation}
Our central result concerns a precise expression for 
\begin{math} \eta  \end{math};  
\par \noindent  
{\bf Theorem:} {\em The information capacity per unit length of a DNA chain 
is given by the thermodynamic expression} 
\begin{equation}
\eta =-\left[\frac{1}{k_B \ln 2}\right]
\left(\frac{\partial \tau }{\partial T}\right)_{\mu_1, \mu_2, \ldots ,\mu_c}.
\label{intro5}
\end{equation}
{\em wherein \begin{math} k_B  \end{math} is Boltzmann's constant}.

The rigorous proof of the theorem will be given in Sec.\ref{st}. 
The only assumptions of the proof reside in the first and 
second laws of statistical thermodynamics. Otherwise, the theorem is 
completely model independent. The importance of a force determination 
of information capacity is that in the laboratory tweezer measurements, 
either optical\cite{Bustamante:2003} or 
magnetic\cite{Leuba:2003}, uniquely determine the DNA chain tension 
\begin{math} \tau (T,\mu_1, \mu_2, \ldots ,\mu_c)  \end{math}. 

In Sec.\ref{ei} the relationship between thermodynamic entropy 
\begin{math} {\cal S}  \end{math} and information 
\begin{math} {\cal H}  \end{math} is reviewed. 
The statistical thermodynamics of long chain molecules is explored  
in Sec.\ref{st} and the proof of the central theorem is provided. 
An order of magnitude statement for 
\begin{math} \eta \end{math} in the human genome is discussed 
in the concluding Sec.\ref{conc}. As one moves up the evolutionary 
scale, the total information capacity in the DNA molecule appears to 
increase.

\section{Entropy and Information \label{ei}}

The connection between entropy and information in statistical thermodynamics 
is well understood\cite{Landau:1999}. The number of microscopic states 
\begin{math} \Omega  \end{math} of a macroscopic system may be written as 
\begin{equation}
\Omega=e^{{\cal S}/k_B}=2^{\cal H},
\label{ei1}
\end{equation}
wherein the thermodynamic entropy \begin{math} {\cal S}  \end{math} is 
determined by 
\begin{eqnarray}
{\cal S}=k_B \ln \Omega \ \ {\rm where} \ \ 
k_B\approx 1.38065\times 10^{-16}\ \ \frac{\rm erg}{\rm ^oK} \ .
\label{ei2}
\end{eqnarray}
The information capacity measured in bits\cite{Adami:2004} is thereby 
\begin{equation}
{\cal H}={\rm lg}\ \Omega = \frac{\ln \Omega}{\ln 2} =\frac{\cal S}{k_B \ln 2}\ . 
\label{ei3}
\end{equation}
where \begin{math} \ln \equiv \log_{\rm e}  \end{math} and 
\begin{math} {\rm lg} \equiv \log_{\rm 2}  \end{math}. 
A discussion of the thermodynamic entropy of a DNA chain follows.

\section{Statistical Thermodynamics \label{st}}

If \begin{math} {\cal E}  \end{math} denotes the energy of a molecular chain 
of length \begin{math} L \end{math} and molecular composition numbers 
\begin{math} {\cal N}_1,{\cal N}_2,\ldots ,{\cal N}_c  \end{math}, then the first 
and second thermodynamic laws for quasi static processes read 
\begin{equation}
d{\cal E}=Td{\cal S}+\tau dL+\sum_{j=1}^c \mu_j d{\cal N}_j\ .
\label{st1}
\end{equation}
The DNA chain quantities 
\begin{math}
({\cal E}, {\cal S}, L, {\cal N}_1,{\cal N}_2,\ldots ,{\cal N}_c )
\end{math}
are all extensive.

Employing extensive scaling 
\begin{equation}
\lambda {\cal E}=
{\cal E}(\lambda {\cal S}, \lambda L, 
\lambda{\cal N}_1,\lambda {\cal N}_2,\ldots ,\lambda{\cal N}_c ),
\label{eq1}
\end{equation}
one finds the Euler equation 
\begin{equation}
{\cal E}={\cal S}\frac{\partial {\cal E}}{\partial {\cal S}}
+L\frac{\partial {\cal E}}{\partial L}+
\sum_{j=1}^c {\cal N}_j\frac{\partial {\cal E}}{\partial {\cal N}_j}\ . 
\label{eq2}
\end{equation}
Eqs.(\ref{st1}) and (\ref{eq2}) imply 
\begin{equation}
{\cal E}=T{\cal S}+\tau L+\sum_{j=1}^c \mu_j {\cal N}_j\ .
\label{eq3}
\end{equation}
Taking the differential of Eq.(\ref{eq3}) and comparing the result to 
Eqs.(\ref{st1}) yields 
\begin{equation}
{\cal S}dT+L d\tau +\sum_{j=1}^c  {\cal N}_j d\mu_j =0 .
\label{eq4}
\end{equation}
Defining the entropy per unit length \begin{math} \sigma \end{math} 
and the molecular densities per unit length 
\begin{math} (\Gamma_1, \Gamma_2,\ldots ,\Gamma_c) \end{math} by 
\begin{equation}
\sigma =\lim_{L\to \infty} \frac{\cal S}{L}\ \ \ {\rm and}
\ \ \ \Gamma_j = \lim_{L\to \infty} \frac{{\cal N}_j}{L} 
\label{eq5}
\end{equation}
together with Eq.(\ref{eq4}) yields 
\begin{equation}
d\tau = -\sigma dT-\sum_{j=1}^c \Gamma_j d\mu_j \ .
\label{eq6}
\end{equation}
The entropy per unit length is thereby 
\begin{equation}
\sigma = -\left(\frac{\partial \tau }{\partial T}\right)
_{\mu_1, \mu_2, \ldots ,\mu_c} = k_B \eta \ln 2   
\label{eq7}
\end{equation}
allowing for the verification of our central theorem. 
Eqs.(\ref{intro3}), (\ref{ei3}), (\ref{eq5}) 
and (\ref{eq7}) yield the required proof of Eq.(\ref{intro5}).

\section{Conclusion \label{conc}}

In order to apply the theorem Eq.(\ref{intro5}), one has to fix the 
molecular chemical potentials. These chemical potentials depend 
on the solution properties of the environment in which the DNA molecule 
resides. Changing these environmental parameters also changes the 
information capacity per unit length of the DNA molecule. We here stress that 
the thermodynamic rule includes the total information capacity of all the 
possible biophysical forms, e.g. four letter coding, ``junk'' DNA 
insertions as well as semiconducting electrons existing in ordered 
water shells coating the DNA chain.
Typical values for the human genome are of order 
\begin{math}
\eta \sim {\rm 10\ byte}/{\rm nanometer}.
\end{math} 
This is completely consistent with Eq.(\ref{intro2}). 
The total information in a DNA molecule gets larger 
as one moves up the evolutionary scale. 
We conclude that the currently understood DNA code is about a
100 megabyte program written on a molecule with about  
10 gigabyte of memory capacity. Most of the 
programing code is beyond our understanding.

\end{document}